\documentclass[12pt]{article}
\usepackage[utf8]{inputenc}
\usepackage{hyperref} 

\usepackage[a4paper, total={6in, 8in}]{geometry}

\newtheorem{definition}{Definition}

\title{Christmas Jump in LIBOR}
\author{Vikenty Mikheev and Serge E. Miheev}
\date{August 2019}

\begin{document}
{
\maketitle
\begin{abstract}
 A short-term pattern in LIBOR dynamics was discovered. Namely, 2-month LIBOR experiences a jump after Xmas. The sign and size of the jump depends on data trend on 21 days before Xmas. 
\end{abstract}

\section{Introduction}

In 1986, a new benchmark interest rate was introduced and named London Interbank Offered Rate (LIBOR). At LIBOR major banks of the world lend to one another in the international interbank market for short-term loans. From mathematical point of view, LIBOR is a sequence of daily changing real values.

LIBOR plays crucial role in Swap Market, where people exchange their loan interests and  can win or lose money depending on their right or wrong predictions of LIBOR dynamics. For example, a person P got a one-million-dollar loan with 5$\%$ interest and a person E borrowed the same amount but with the interest $2\% + $LIBOR. After some time they decide to exchange their interest rates because P thinks that LIBOR will go lower than $3\%$ but E hopes that it will go higher than $3\%$. Both their opinions are based on some prediction methods, even if it is just an intuition. We intend to bring another prediction tool into the game. A curious reader may find more complex models and measures on LIBOR for different problems  \cite{Jamshidian1997LIBORAS}, \cite{Schoenmakers2005Robust}, \cite{Moreni} and
\cite{Hinch}.

Thus, here we are not interested to LIBOR nature per se but to its volatility only.  More precisely, we study the behavior of LIBOR after Christmas from December 26 to December 31. 

So, how does Christmas affect dynamics of LIBOR until the next holiday?

\section{Definition and Models}

The research we conducted indicates convincingly that a {\it jump} is present. But what is a jump in a discrete sequence of numbers? The following seems to be the most acceptable

\begin{definition}
\label{Def:Jump}
We introduce two approximations for discrete data. One for a certain period before considered date (in this case Dec 25) and the other one for a certain period after the date. Let them accordingly have forms: $B_0+A_0(x)$ and $B_1+A_1(x)$,  where $x$ is a date and $A_0, A_1$ are continuous functions. Moreover, $A_0(Dec 25)=A_1(Dec 25)=0$. The difference $B_0-B_1$ is the \underline{jump}.
\end{definition}

It is easy to see that the so-defined jump depends on the type of approximation, i.e. on how the functions $A_0, A_1$ are selected; and on the amount of the input data. We still have to decide the amount of input data to calculate the approximating pair $B_0, A_0$. Notice that the amount of the data for $A_1$ and $B_1$ is three pairs (date, LIBOR of this date) only because there are exactly three working banking days between Xmas and NYE. Data source is available at \cite{dataLIBOR} or multiple other sources.

Variability of the data due to random factors leads to the choice of the simplest approximation. We use linear approximating functions, which coefficients may be found by linear regression.
We restrict ourselves to LIBOR data for the last 22 years, because it is natural to expect the evolution of LIBOR behavior over the years.

So, for some year $j$ 
in some set $J$ taken sequentially  with no gaps
from $\{$1997, ..., 2018$\}$ data are taken for
15 banking days $x'_{-15}, ..., x'_{-1}$ (corresponding to 21 calendar days) preceding Xmas of year $j$. As all the days are in December, for simplicity of following constructions we may decrease them by 25 without problem of passing days to another month.
So, $x_i:=x'_i-25,\ \ i=-15,...,-1$.
Each $x_i$ corresponds to $y_i$, which is the annual interest rate of LIBOR for 2 months on day $x_i$. Using them we build a linear regression 
\begin{equation}
\label{RegrOnYearBeforXmas}
\hat{y}(x)=a_jx+b_j,
\end{equation}
\noindent
or
$$y_i =  \hat{y}(x) + error_i,$$ 
 where $x$ is a December day minus 25.

That is, in terms of Definition \ref{Def:Jump}, $B_0: = b_j, A_0 (x): = a_j x$.
Since the current trend of LIBOR (meaning the rate of growth or decrease) does not change a lot over a short time interval, it is almost the same before and after Xmas. Therefore, we seek an approximation after Xmas in the following form: $\hat{y}(x)=a'_j x+b'_j, $
 where $a'_j=a_j$ \ 
 and $x$ is a December day decreased by 25.
 In terms of Definition \ref{Def:Jump}, \ $B_1: = b’_j, \ A_1 (x): = a’_j x\equiv A_0(x)$. Hence, there is only one unknown parameter $B_1$.  It can also be found by linear regression procedure or it can be calculated simply as the average of values $y_i - a_j x_i$, where $i$ runs 
 $1,2,3$ and $x_i\in\{27-25,...,31-25\}=\{2,...,6\}$
(There are exactly 3 bank days between Xmas and New Year.)

Thus, for each selected year $j$ there is a relationship $(b_j,a_j)\rightarrow \hat{b}_j$. According to Definition \ref{Def:Jump} the difference $\Delta_j:=\hat{b}_j - b_j$ is the jump we have been looking for. Having such connections over 22 years, one can try to find a pattern. To do that, we turn to 
linear-quadratic  regression in shorten variant deprived pure square.
This time we approximate on two-dimensional nodes, in other words the approximating function is of the form: 
\begin{equation}
\label{Eq:BigRegression}
F(a,b):=\beta_0+\beta_1a+\beta_2b +\beta_3 ab    
\end{equation} 
with an approximation table $F_J(a_j,b_j)\approx \Delta_j, \ j\in J\subset \{1995,\ldots, 2018\}$. Subindex $J$ at $F$ points at which subset of years over the past 22 has been chosen to construct the regression. The remaining years will be used to verify the statistical reliability of the result.

\section{Calculations and Results}

We conducted the process above for several different numbers of years for $F$ regression (from 5 to 20 years), different LIBOR data (Overnight, 1 month, 2 months etc). The most convincing results were have obtained with the following set ups: 21 calendar day regression for each year from 15-year interval; 2-month-loan values of LIBOR. 

Observe the results in the Table \ref{tab:resInteract}.

\begin{table}[h]
\caption{Model $\Delta \approx \beta_0+\beta_1a+\beta_2b +\beta_3 ab$ and its predictions with p-values $p$ for 2019 year.}
    
    \centering
    \begin{tabular}{|c|c|c|c|c|c|}\hline
      Data   &  2000-14	&	2001-15	&	2002-16	&	2003-17	&	2004 - 2018 \\ \hline
      Pred. on &     2015 &     2016 &     2017 &     2018 &     2019 \\ \hline
       
     $\widehat{\beta_0}$  & -2.96E-3 & 4.7E-4 & 4.1E-4 & 0.00327 & 0.00473 ($p$=0.138)\\
     $\widehat{\beta_1}$  & -9.286 & -9.337 & -9.321 & -9.278 & -9.265 ($p$=1.43E-13)\\
     $\widehat{\beta_2}$ & 1.8E-4 & -0.00209 & -0.00220 & -0.00204 & -0.00238 ($p$=0.088)\\
     $\widehat{\beta_3}$ & 1.91521 & 1.99021 & 1.98676 & 2.02619 & 2.016 ($p$=5.05E-09)\\  \hline \hline
     Adj $R^2$ & 0.98 &		0.99	&	0.99	&	0.99	&	0.99 \\ \hline 
     Pred. val & -0.0709	&	-0.0306	&	-0.0548	&	-0.0180	& Wait for data \\ 
     Real val & -0.0599	&	-0.0269	&	-0.0291	&	-0.0228	&  till 12/24/2019\\  \hline
     $\hat{L}^{mean}$ & 0.5024	&	0.8146 &		1.5967	&	2.6222 &  Wait for data\\
     ${L}^{mean}$ & 0.5134	&	0.8183	&	1.6224 &		2.6174 & till 12/24/2019\\ \hline
     Error & -0.0110	&	-0.0037	&	-0.0257	&	0.0048 &\\	\hline
      \end{tabular}
    \label{tab:resInteract}
\end{table}

So, our prediction for the jump formula after Xmas 2019 are:
\begin{equation}
\label{PredFor2019}
\widehat{\Delta}_{2019} =
0.0048 - 9.2646a_{2019}-0.0024b_{2019} +2.0161a_{2019}b_{2019}
\end{equation}

It may be activated at Dec 24 2019 as following:

At this day extract data 
from \cite{dataLIBOR} for bank days since Dec 21 till Dec 24 (in 2019, of course).
Build 15 pairs $(x_i,y_i), \ \ i=-15,...,-1$. Put them into any program
to find linear regression, for example our code in R can be found in 
\url{https://github.com/keshmish/Chistmas-Jump-in-LIBOR.git}
The result of its work is two numbers: corresponding to free term is $b_{2019}$, the other one is
$a_{2019}$. Substitution them to (\ref{PredFor2019}) yields the jump.

And what is about last Christmas \cite{LiborGone} of LIBOR existence, 2020? At present the data for 2019 year
do not exist. Therefore we can only propose the same coefficients as in (\ref{PredFor2019}).
It can worse the prediction but quite a little. At 2020 Xmas will happened in Tuesday, so Dec 24 also is bank day. So, this day one should do the same that we suggested above for 2019. Namely, take 15 pairs of data $(x_i,y_i), \ i=-15,\ldots, -1$ and use them as input for linear regression procedure with approximating function (\ref{RegrOnYearBeforXmas}) in R, Python or any other language. As a result one obtains two numbers $a_{2020}, b_{2020}$. Then put them in the right hand side of the formula (\ref{PredFor2019}) on the place of $a_{2019}, b_{2019}$, respectively.

In fact, a reader can observe that predictors of (\ref{Eq:BigRegression}) don't change too dramatically with the shift of 15 regression years by one. Thus, the prediction for 2020 with 2004-2018 model may appear to be quite good.

The prediction of the jump can be used to predict the mean LIBOR after Xmas before NYE ($L^{mean}_j$). Let us show some formulas.

According to Definition \ref{Def:Jump} the jump with approximations above is $\Delta_j = \hat{b}_j - b_j$, where $\hat{b}_j = \arg\min_{b}\{\sum_{i=1}^3 (a_jx_i +b - y_i )^2 \}$, which is equivalent to $\hat{b}_j = \frac{1}{3}\sum_{i=1}^3 (y_i-a_j x_i  )$. Hence 
\begin{equation}
\label{Eq:DeltaAve}
\Delta_j = \frac{1}{3}\sum_{i=1}^3(y_i- a_jx_i) - b_j = \frac{1}{3}\sum_1^3y_i -\frac{1}{3}\sum_{i=1}^3(a_jx_i+b_j).
\end{equation}

Notice that the last term of (\ref{Eq:DeltaAve}) is nothing but predicted after Christmas mean value of LIBOR ($\hat{L}^{mean}_j$) according to the regression for $j$-th year. Thus, 
$$
L^{mean}_j=\hat{L}^{mean}_j + \Delta_j = \hat{L}^{mean}_j + \hat{\Delta}_j + (\Delta_j - \hat{\Delta}_j).  
$$
If as estimate of $L^{mean}_j$ we take $\hat{L}^{mean}_j + \hat{\Delta}_j$, then  its absolute error equals to $  \hat{\Delta}_j - \Delta_j$. The latter difference according to our calculations for years 2015, 2016, 2017, 2018 was always less by absolute value than $| \hat{\Delta}_j|$.

\section{Conclusion}

We have found a short-term pattern in LIBOR dynamics. Namely, 2-month LIBOR experiences a jump after Xmas. The sign and size of the jump depends on data trend on 21 days before Xmas. The results are obtained in the form of the jump per se and as mean predicted value of LIBOR between Xmas and NYE. A swap market player may try to use this information to predict behaviour LIBOR to do a better game on his part. 
For Xmas of 2019 one on a date of Dec 24 can compute $a$ and $b$ according to (\ref{RegrOnYearBeforXmas}) on 21 calendar days and use the formula (\ref{PredFor2019}) to predict the jump after Xmas.

\section{Acknowledgements}
Institute for Mathematics and its Applications (\url{https://ima.umn.edu/node}), U. of Minnesota, Fadil Santosa, Daniel Spirn, Davood Damircheli, Anthony Nguyen, Samantha Pinella. 
}

 	\bibliographystyle{alpha}
\end{document}